\documentclass[letterpaper,preprintnumbers,superscriptaddress,nofootinbib,number]{revtex4}
\usepackage{graphicx}
\usepackage{color}
\usepackage{bm}
\usepackage{amsmath,amssymb,url,hyperref}
\newcommand\ignore[1]{}

\begin{document}

\preprint{JLAB-THY-12-1591}
\title{Nuclear correlation functions in lattice QCD } \author{William
  Detmold} \author{Kostas Orginos} \affiliation{Department of Physics,
  The College of William \& Mary, Williamsburg, VA 23187, USA}
\affiliation{Jefferson Lab, Newport News, VA 23606, USA }

\begin{abstract}
  We consider the problem of calculating the large number of Wick
  contractions necessary to compute states with the quantum numbers of
  many baryons in lattice QCD. We consider a constructive approach and
  a determinant-based approach and show that these methods allow the
  required contractions to be performed 
  in computationally manageable amount of time
   for certain choices of
  interpolating operators. Examples of correlation functions computed
  using these techniques are shown for the quantum numbers of the
  light nuclei, $^4$He, $^8$Be, $^{12}$C, $^{16}$O and $^{28}$Si.
\end{abstract}
\maketitle

\section{Introduction}

The ab initio approach to nuclear physics from the underlying theory
of the strong interactions, Quantum Chromodynamics (QCD), is hampered
by the many body nature of the nuclear problem. In principle, QCD and
the electroweak interactions give rise to all the rich and complex
phenomena of nuclear physics, yet it is only recently that the first
QCD studies of multi-baryon systems have appeared \cite{Beane:2009gs,
  Yamazaki:2009ua,Beane:2010hg,Beane:2011xf,Inoue:2011tk,Beane:2011iw,
  Yamazaki:2011nd,NPLQCD_hyper}.  The reason for this is
twofold. Firstly, the Monte-Carlo evaluation of correlation functions
of multi-baryon systems converges slowly, requiring a large number of
measurements before the necessary precision is reached (this issue will not be addressed here).
 Secondly,
systems with the quantum numbers of many nucleons and hyperons are
complex many-body systems with complicated spectra and there are a
multitude of physically relevant states that can be studied in QCD.
Even for a given set of quantum numbers, additional complexity appears
at the quark level; the number of Wick contractions required to
construct systems for large atomic number grows factorially, scaling
as $n_u!n_d!n_s!$ where $n_{u,d,s}$ are the numbers of up, down, and
strange quarks required to construct the quantum numbers of the state
in question. In many situations, this is a naive counting as there are
many cancellations and contributions that are identical.  However,
the a priori identification of these simplifications is a non-trivial
task.  In addition to the problem of Wick contractions, the number of
terms in the interpolating fields of multi-nucleon systems also
typically grows exponentially with the size of the system. This
potentially more serious problem is similar in nature to the problem
of the exponential growth of nuclear wave-functions faced in nuclear
structure calculations where phenomenological potential models
describing the low energy nucleon-nucleon interactions are used.

In this paper, we present a systematic method for the construction of
nuclear interpolating fields for multi-baryon systems in lattice QCD
(LQCD) (see Ref.~\cite{Doi:2012xd} for related work). We demonstrate
that the Grassmannian nature of the quark fields can be used to our
advantage, in some cases resulting in particularly simple nuclear
interpolating fields.  In addition, we present two approaches that
ameliorate the cost of contractions, the most efficient of which
scales only polynomially in the number of quarks involved in the
contraction. Using these methods we compute LQCD correlation functions
with the quantum numbers of the light nuclei, $^4$He, $^8$Be,
$^{12}$C, $^{16}$O and $^{28}$Si, demonstrating that correlation
functions relevant to the study of nuclei in QCD can be constructed.

\section{Nuclear Interpolating fields}

In order to calculate nuclear correlation functions, we first need to
construct quark level nuclear interpolating fields. This is, in
principle, straightforward and, in practice, it resembles the
construction of quark model wave-functions for
baryons~\cite{Feynman:1971wr}.  A general quark-level nuclear
interpolating field with atomic number $A$ containing $n_q=3A$ quarks
has the form
\begin{equation}
  \label{eq:gen_nuc}
  \bar{\cal N}^{h}= \sum_{\bf a} w^{a_1,a_2\cdots a_{n_q}}_h \bar{q}(a_1) \bar q(a_2)\cdots \bar q(a_{n_q}) \,,
\end{equation}
where the $\bar q_{a_i}$ are the quark fields, the $a_i$ are  generic
indices which combine the colour, spinor, flavour, and spatial indices of
the quark and ${\bf a}$ is a compound index representing the
$n_q$-plet $a_1,a_2\cdots a_{n_q}$.  Given that calculations are
performed on a discrete lattice, the spatial degrees of freedom are
finite and countable, and as a result we can use an integer index to
describe them.  Here the quark fields are all at the same time $t$.
The index $h$ on the nuclear interpolating field is a set of quantum
numbers that identify the nuclear state, including its momentum,
angular momentum, isospin and strangeness. The Grassmannian nature of
the quark field dictates that the tensor $w^{a_1,a_2\cdots a_{n_q}}_h$ is
totally antisymmetric under the exchange of any two indices.  If the indices $a_i$ can have a total of $N$
possible values, then, ignoring the detailed flavour structure, the
total number of non-vanishing terms in the above sum is
\begin{equation}
  \label{eq:Nterms}
  \frac{N!}{(N-n_q)!} \,.
\end{equation}
However, many of these terms correspond to permutations of the quark
fields, and the total number of unique terms (terms that are not a
permutation of any other term) is
\begin{equation}
  \label{eq:NtermsTrue}
  \frac{N!}{n_q! (N-n_q)!} \,.
\end{equation}
For a generic spatial structure of the interpolating field, $N$
corresponds to the total number of quark degrees of freedom on a
time-slice ($N=12L^3$, where $L$ is the spatial dimension of the
lattice) and one may be discouraged by the feasibility of the task of
building quark level nuclear interpolating fields. However, a number
of simplifying factors are omitted in the above discussion.  As we
consider interpolating fields with definite transformation properties
under the symmetries of QCD, large numbers of terms in the nuclear
interpolating field vanish.  The first major reduction comes from the
fact that only colour singlets need to be considered.  In addition,
considering only interpolating fields of definite parity, angular
momentum~\footnote{For simplicity, we refer to the irreducible
  representation of the lattice symmetry group as angular momentum.},
isospin and strangeness, forces several elements of the tensor
$w^{a_1,a_2\cdots a_{n_q}}_h$ to vanish. Finally, the most drastic
reduction of the non-zero tensor elements can be achieved using simple
spatial wave-functions.  At this time, having recognised that only a
small fraction of the terms in the sum of Eq.~\ref{eq:gen_nuc} are
non-zero, as well as the fact that the tensor $w^{a_1,a_2\cdots
  a_{n_q}}_h$ is totally anti-symmetric, we can introduce the reduced
weights $\tilde{w}^{(a_1,a_2\cdots a_{n_q}),k}_h$ 
which are the minimal set of non-zero numbers required to completely describe the interpolating field.
The $n_q$-plet $(a_1,a_2\cdots
a_{n_q})$, is an ordered list of indices that represents a class of
terms in Eq.~\ref{eq:gen_nuc} that are all permutations of each
other. The index $k$ on the reduced weights enumerates the number of
classes that the tensor $w^{a_1,a_2\cdots a_{n_q}}_h$ decomposes into.
With these reduced weights, Eq.~\ref{eq:gen_nuc} can be re-written as
\begin{equation}
  \label{eq:gen_nuc_red}
  \bar{\cal N}^{h}= \sum_{k=1}^{N_w} \tilde{w}^{(a_1,a_2\cdots a_{n_q}),k}_h 
  \sum_{\bf i} \epsilon^{i_1,i_2,\cdots,i_{n_q}}
  \bar{q}(a_{i_1}) \bar q(a_{i_2})\cdots \bar q(a_{i_{n_q}}) \,,
\end{equation}
where $N_w$ is the total number of reduced weights, ${\bf i}$
represents the $n_q$-plet $(i_1,i_2\cdots i_{n_q})$ and
$\epsilon^{i_1,i_2,\cdots,i_{n_q}}$ is a totally anti-symmetric tensor
of rank $n_q$ with
\begin{equation}
  \epsilon^{1,2,3,4,\cdots,{n_q}} = 1\,.
  \nonumber
\end{equation}
The above expression is the simplest form of the quark-level nuclear
interpolating field and is completely described by the reduced
weights.  As an example, using a single point spatial wave function,
the numbers of terms contained in the simplest interpolating fields
for the proton, deuteron, $^3$He and $^4$He, are $N_w=9,\, 21,\, 9$,
and 1, respectively.

\subsection{Hadronic Interpolating Fields}
Having now written down a general nuclear interpolating field with
quantum numbers $h$, we need to calculate the reduced weights
$\tilde{w}^{(a_1,a_2\cdots a_{n_q}),k}_h$ in an efficient manner.  In
principle, this can be achieved directly from quark fields by imposing
the desired transformation properties.  However, in certain cases, it
is advantageous to proceed by first constructing hadronic
interpolating fields from which the quark interpolating fields are
derived.

The hadronic interpolating fields assume a form analogous to that of
the quark interpolating fields.  The baryons that make up the nucleus
are also fermions, hence the general structure outlined above can be
directly transcribed here. In terms of baryons, a nuclear
interpolating field of a nucleus of atomic number $A$ is
\begin{equation}
  \label{eq:gen_nuc_bar}
  \bar{\cal N}^{h}= \sum_{k=1}^{M_w} \tilde{W}^{(b_1,b_2\cdots b_{A})}_h 
  \sum_{\bf i} \epsilon^{i_1,i_2,\cdots,i_{A}}
  \bar{B}(b_{i_1}) \bar B(b_{i_2})\cdots \bar B(b_{i_{A}}) \,,
\end{equation}
where $M_w$ is the number of hadronic reduced weights
$\tilde{W}^{(b_1,b_2\cdots b_{A})}_h$, $B(b_i)$ are baryon
interpolating fields and the $b_i$ are generic indices that includes
parity, angular momentum, isospin, strangeness, and spatial indices.
Unlike the quark fields which are fundamental degrees of freedom, the
baryon interpolating fields are composite objects, hence there is a
large number of such interpolating fields for a given set of quantum
numbers.  For simplicity, as well as efficiency of the resulting
nuclear interpolating fields, we will use a single interpolating field
per baryon, selected so that it has good overlap with the single
baryon ground state, as well as being comprised of a small number of quark
level terms.
The utility of the above form of the nuclear interpolating fields is
twofold. Firstly, it allows us to derive the reduced weights we need for
Eq.~\ref{eq:gen_nuc_red}. Secondly, interpolating fields that are
derived starting from Eq.~\ref{eq:gen_nuc_bar} may have better overlap
with the nuclear ground states as it is well-known that hadronic
degrees of freedom provide a successful description of much of nuclear
physics.

The calculation of the reduced weights, $\tilde{W}^{(b_1,b_2\cdots
  b_{A})}_h$, in the hadronic interpolating field is
straightforward. It amounts to combining individual hadrons of given
quantum numbers to build a multi-hadron state of definite parity,
angular momentum, isospin, and strangeness. This construction can be
readily automated and can be performed recursively using the known
Clebsch-Gordan coefficients of SU(2) for both the spin and isospin (or
SU(3) flavour if so desired).  In principle, one can use all the octet
and decuplet baryons in Eq.~\ref{eq:gen_nuc_bar}, however, for most
practical purposes, restricting to the positive parity octet baryons
is sufficient.  For example, for $A=2$, $I=J=0$, $S=-2$, if we
restrict the spatial wave-function to single point, there are three
simple hadronic interpolating fields
 $$\Lambda^\uparrow\Lambda^\downarrow\,,$$
$$\frac{1}{\sqrt{3}}\left[\Sigma^{+\uparrow}\Sigma^{-\downarrow}-
  \Sigma^{0\uparrow}\Sigma^{0\downarrow}+
  \Sigma^{-\uparrow}\Sigma^{+\downarrow}\right]\,,$$ and,
$$\frac{1}{2}\left[\Xi^{0\uparrow}n^{\downarrow} -\Xi^{-\uparrow}p^\downarrow -
  \Xi^{0\downarrow}n^\uparrow +
  \Xi^{-\downarrow}p^\uparrow\right]\,,$$ where $B^\uparrow$ and
$B^\downarrow$ represent the spin up and down polarisations of the
baryon, $B$, respectively.  In this example, the reduced weights can
be directly read off from these equations.

We have written a {\tt c++} symbolic manipulation program that
generates the hadronic reduced weights using the above approach. In
Ref.~\cite{NPLQCD_hyper}, we have used this to produce a complete
basis of orthonormal interpolating fields with spatial wave-functions
restricted to a single point for all nuclei up to $A=4$ and have also
constructed a selection of states for $A>4$.  Generically for larger
$A$, more complicated spatial wave-functions are required because of
the Pauli exclusion principle, resulting in an exponential growth of
the number of possible interpolating fields as $A$ increases (this
reflects the problem faced in nuclear structure calculations  as $A$
becomes large).  In certain cases, the Grassmannian nature of the
quark fields is also advantageous, drastically reducing the number of
non-zero reduced weights.  Making use of this feature, we have been
able to find particularly simple wave-functions for systems as large
as $A=28$.

\subsection{Quark Interpolating Fields}
 
The reduced weights of the quark interpolating fields of
Eq.~\ref{eq:gen_nuc_red} can be calculated by equating the two forms
of the nuclear interpolating fields
\begin{eqnarray}
  \bar{\cal N}^{h}&=& \sum_{k=1}^{M_w} \tilde{W}^{(b_1,b_2\cdots b_{A})}_h 
  \sum_{\bf i} \epsilon^{i_1,i_2,\cdots,i_{A}}
  \bar{B}(b_{i_1}) \bar B(b_{i_2})\cdots \bar B(b_{i_{A}}) \nonumber\\
  &=& 
  \sum_{k=1}^{N_w} \tilde{w}^{(a_1,a_2\cdots a_{n_q}),k}_h 
  \sum_{\bf i} \epsilon^{i_1,i_2,\cdots,i_{n_q}}
  \bar{q}(a_{i_1}) \bar q(a_{i_2})\cdots \bar q(a_{i_{n_q}})
  \,,
  \label{eq:had_to_quark}
\end{eqnarray} 
and replacing the baryon objects by their quark interpolating fields.
A single baryon interpolating field is written in terms of quark
fields as
\begin{equation}
  \bar{B}(b)=  \sum_{k=1}^{N_{B(b)}} \tilde{w}^{(a_1,a_2,a_3),k}_b 
  \sum_{\bf i} \epsilon^{i_1,i_2,i_3}
  \bar{q}(a_{i_1}) \bar q(a_{i_2})\bar q(a_{i_3})\,,
\end{equation}
where $N_{B(b)}$ is the number of terms in the single baryon $B(b)$
interpolating field.  For single baryon interpolating fields, the
weights, $\tilde{w}^{(a_1,a_2,a_3),k}_b$, have been presented
in~\cite{Basak:2005ir} (the colour factors necessary for our
formulation are not included in Ref.~\cite{Basak:2005ir} but can be
trivially added).  The process of deriving the reduced weights
$\tilde{w}^{(a_1,a_2\cdots a_{n_q}),k}_h$ from
Eq.~\ref{eq:had_to_quark}, can be automated and we perform it within
our symbolic manipulation program.  An interesting feature that arises
from the calculation of the reduced weights $\tilde{w}^{(a_1,a_2\cdots
  a_{n_q}),k}_h$ is that if we restrict ourselves to simple spatial
wave-functions making use of only few spatial points, then the
expected exponential growth of the number of terms in the nuclear
interpolating field is eliminated. A careful selection of the spatial
wave-functions used can eliminate this problem, in principle for
arbitrarily large nuclei. However, restriction to a small number of
quark degrees of freedom also makes it impossible to construct certain
states (an example is presented in Ref.~\cite{NPLQCD_hyper} where the
two baryon symmetric flavour octet was found to be inaccessible).

\section{Techniques for multi-baryon contractions}

In this section, we consider how the interpolating fields constructed
in the previous section can be used to generate the correlation
functions of multi-baryon systems. A general multi-hadron two point
function is given by
\begin{equation}
  \langle {\cal N}^h_1(t)\bar{\cal N}^{h}_2(0)\rangle = \frac{1}{\cal Z} \int {\cal D} {\cal U} {\cal D} q {\cal D} \bar q \;
  {\cal N}^{h}_1(t) \bar{\cal N}^{h}_2(0) \; e^{-{\cal S}_{QCD}}\,,
  \label{eq:2pt}
\end{equation}
where ${\cal S}_{QCD}$ and $\cal Z$ are the QCD action and partition
function respectively, and ${\cal DU}$, ${\cal D} q {\cal D} \bar q $ are the gluon and quark field integration
measures respectively.  We have also introduced explicit dependence of the
interpolating fields on the Euclidean time separation, $t$, and
consider a two point function with different creation and annihilation
interpolating fields with commensurate quantum numbers.  For a given
choice of the interpolating fields, it is straightforward to perform
the Grassmann integral over the quark fields and re-write the
correlation function in terms of the quark propagators. However, for
an efficient calculation of the two point function we need to be
mindful of the structure of the interpolating fields.

One successful class of interpolating fields for two or more hadron
systems is one in which a plane wave basis at the level of the hadronic
interpolating fields is used. This amounts to  projecting the individual hadrons comprising the
multi-body system to definite momentum
states,
while preserving the spatial transformation properties of the overall multi-hadron system~\cite{Beane:2005rj,Beane:2006mx,Beane:2007es,Detmold:2008fn,Beane:2009kya,Beane:2009py,Beane:2009gs,Beane:2011sc}.
In this case, the complexity of the spatial wave-function is such that
the number of terms contributing to Eq.~\ref{eq:gen_nuc_red} is rather
large and hadronic interpolating fields have to be used in order to
build the desired two point function. Constructing these types of
interpolating fields both at the source and the sink becomes
computationally expensive because a large number of quark propagators
that are required. Nevertheless, this method has been employed for
meson-meson and multi-meson
spectroscopy~\cite{Beane:2011sc,Shi:2011mr,Detmold:2012wc,Dudek:2012gj}.
For the case of multi-meson systems, special contraction methods were
required~\cite{Detmold:2010au,Shi:2011mr,Detmold:2012wc}. For
multi-baryon systems, the problem is more complex and will be the
subject of further investigations.
A further approach is to consider correlation functions in which the
quark creation interpolating fields (source) have simple spatial
wave-functions with few degrees of freedom (for example, restricted to
a few spatial locations), while using a plane wave basis for the
hadronic interpolating fields at the sink. Finally, as we shall
discuss below, sufficiently simple nuclear interpolating fields exist,
where the number of terms contributing in Eq.~\ref{eq:gen_nuc_red} is
small and factorization into hadrons is not computationally necessary.

\subsection{Hadronic blocks}

The quark propagator from a single source point, $x_0=({\bf x}_0,0)$,
can be used to construct baryon building blocks with quantum numbers
$b$ and momentum ${\bf p}$, as:
\begin{equation}
  \label{eq:1}
  {\cal B}_b^{a_1,a_2,a_3}({\bf p},t;x_0)= \sum_{\bf x}e^{i{\bf p}\cdot{\bf
      x}} 
  \sum_{k=1}^{N_{B(b)}} \tilde{w}^{(c_1,c_2,c_3),k}_b 
  \sum_{\bf i} \epsilon^{i_1,i_2,i_3}
  S(c_{i_1},x;a_1,x_0)
  S(c_{i_2},x;a_2,x_0)
  S(c_{i_3},x;a_3,x_0)
  \,,
\end{equation}
where $S(c,{\bf x},t;a,x_0,0)$ is the quark propagator from $x_0$ to
$x=({\bf x},t)$ and $c_i$, $a_i$ are the remaining combined
spin-colour-flavour indices.  In this notation, the sink indices are
kept to the left of the source indices and the spatial indices are
displayed explicitly as they play an essential role in the
construction of the block.  This baryon block corresponds to the
propagation of an arbitrary three-quark state from the source to the sink
where it is annihilated by the prescribed baryon interpolating field.
As discussed above, we have chosen to momentum project these blocks at
the sink to a given momentum ${\bf p}$ to allow control of the total
momentum of multi-hadron systems, although this is not necessary and
other forms of blocks can be envisaged.

We can generalise these blocks to allow the quark propagators to
originate from different source locations, $x_0^{(1)},
x_0^{(2)},\ldots$, as necessary, using
\begin{equation}
  \label{eq:6}
  {\cal B}_b^{a_1,a_2,a_3}({\bf p},t;s_1,s_2,s_3)= \sum_{\bf x}e^{i{\bf p}\cdot{\bf
      x}} 
  \sum_{k=1}^{N_{B(b)}} \tilde{w}^{(c_1,c_2,c_3),k}_b 
  \sum_{\bf i} \epsilon^{i_1,i_2,i_3}
  S(c_{i_1},{\bf x};a_1,x_0^{(s_1)})
  S(c_{i_2},{\bf x};a_2,x_0^{(s_2)})
  S(c_{i_3},{\bf x};a_3,x_0^{(s_3)})
  \,,
\end{equation}
where the $x_0^{(k)}$ label the source locations. These blocks can be
further generalised to allow for non-trivial single hadron spatial
wave-function at the sink, but we will not consider this case
further. It may also be advantageous to consider more complicated
multi-hadron blocks similar to those implemented in
Ref.~\cite{Yamazaki:2009ua} although the storage requirements grow
rapidly with number of baryons in the block.

\subsection{Quark-hadron contractions}
\label{sec:quark-hadr-contr}

\begin{figure}[!h]
  \centering
  \includegraphics[width=5cm]{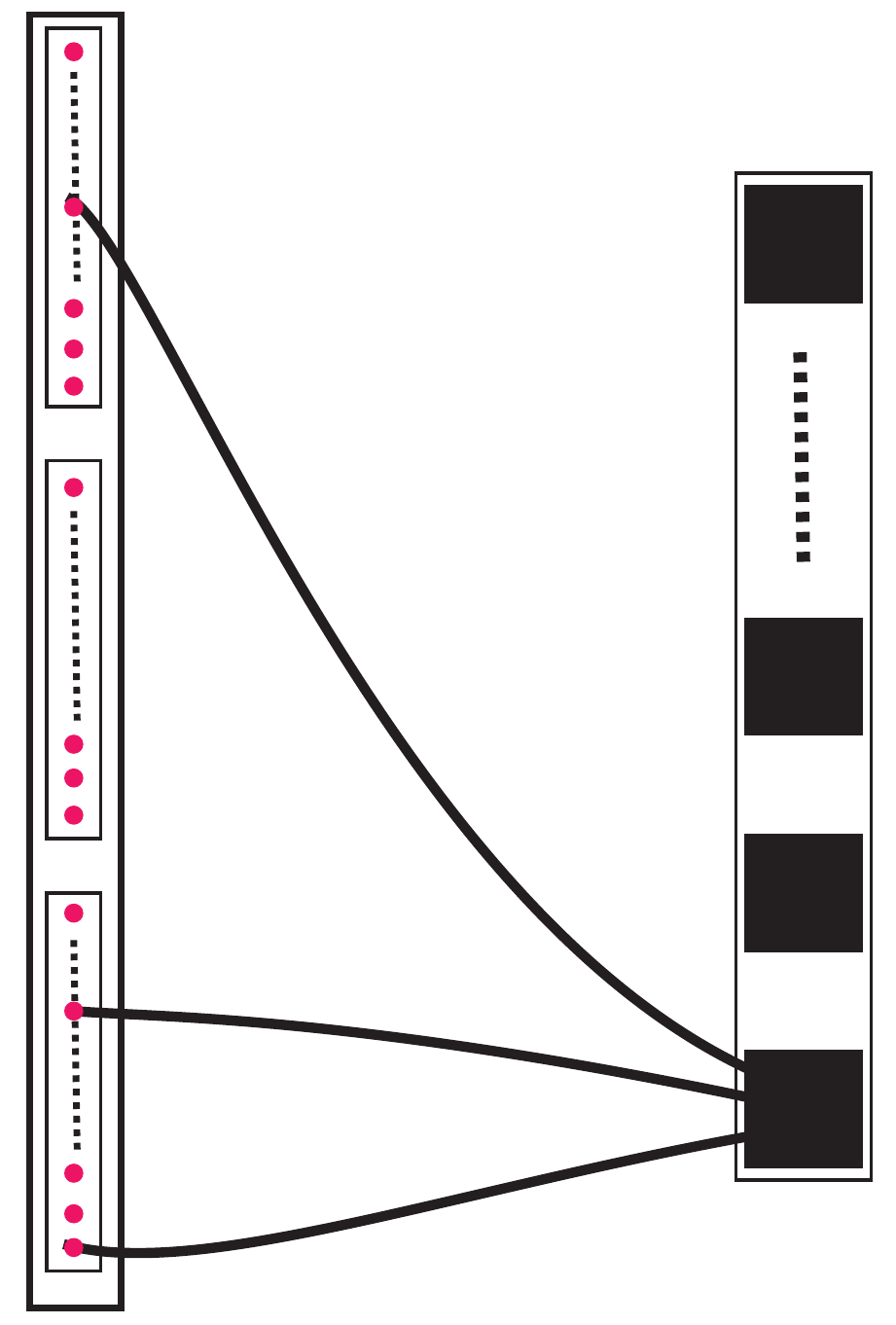}\qquad\qquad\qquad\qquad\qquad
  \includegraphics[width=5cm]{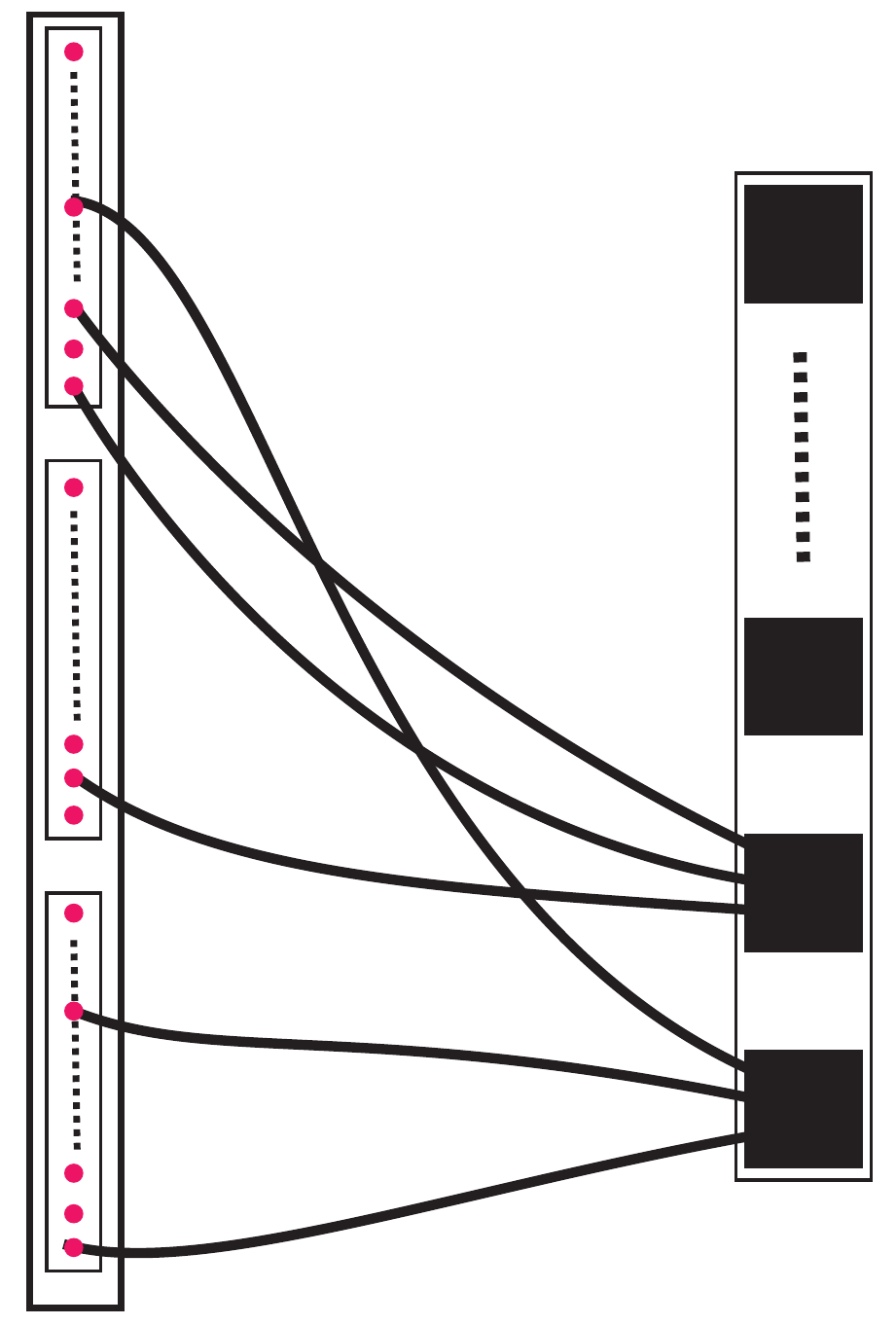} \\
  (a)\hspace*{8cm}(b)
  \caption{Illustration of steps one and two of the quark--hadron
    contraction method. The small circles in the left hand of the
    figures correspond to the quarks in the source interpolating field
    while the large squares and lines extending from them correspond
    to the hadronic blocks.}
  \label{fig:q-h}
\end{figure}

Using the building blocks described above, we can consider correlation
functions in which quark level interpolating fields are used at the
source and their hadronic counterparts are used at the sink.  The
contractions are performed by iterating over all combinations of
source and sink interpolating field terms and connecting the source
and sink with the appropriate sets of quark propagators. For a given
pair of source and sink interpolating field terms, this amounts to
selecting the components dictated by the source quark interpolating
field from the product of blocks dictated by the hadronic sink
interpolating field.  The Wick contractions are implemented by
performing this selection in all possible ways.  This proceeds by
taking the first hadron in the hadronic wave-function at the sink,
replacing it by the appropriate hadron block and selecting the three
free indices in all possible ways from the pool of indices dictated by
the source quark interpolating field, keeping track of the appropriate
permutation sign.  Following this, the second baryon component in the
hadronic (sink) interpolating field term is replaced with the
appropriate block and the free indices are contracted with the
remaining free indices in the source quark interpolating field term in
all possible ways. These first steps are illustrated in
Fig.~\ref{fig:q-h} and the procedure continues until all hadrons in
the sink interpolating field term have been contracted, necessarily
using all available quark indices at the source.
The result is then multiplied by the weights of the source and sink
terms under consideration and added to the correlation function.  The
contraction is complete after all combinations of source and sink
interpolating field terms have been considered. The process described
here is independent of the the source and sink interpolating fields
and can be applied to any correlation function.  Further reductions of
the total cost of the algorithm may be possible by studying the
symmetry properties of a particular pair of source-sink interpolating
fields.  However, such reductions are not generic, hence we do not
consider them further.  The procedure described has been used to
perform the contractions needed for the large class of interpolating
fields considered in the study of the spectrum of hyper-nuclei up to
$A=5$ in Ref.~\cite{KOtalk,NPLQCD_hyper}.

For large numbers of baryons ($A>8$ for protons and neutrons alone), it
is necessary to use multiple source locations because of the Pauli
exclusion principle. In this case, the generalised blocks in
Eq.~\ref{eq:6} can be used with the algorithm presented above.

\subsection{Scaling}

From the above description, it is clear that this algorithm will in
general scale as
\begin{equation}
  M_w \cdot N_w  \cdot  \frac{ (3 A) !}{(3!)^A}\,,
\end{equation} 
where $A$ is the atomic number and $M_w$ and $N_w$ are the number of
terms in the sink and source interpolating fields respectively. In
addition, the fact that the hadron blocks are completely
anti-symmetric under all quark exchanges has been taken into
account. If we also take into account that the strong interactions are
flavour-blind and consider only octet baryon building blocks, this
reduces to
\begin{equation}
  M_w \cdot N_w  \frac{ n_{u} ! n_d !  n_s!}{2^{A-n_{\Sigma^0} - n_\Lambda}}\,,
\end{equation} 
where $n_{\Sigma^0}$ and $ n_\Lambda$ are the number of $\Sigma^0$ and
$\Lambda$ baryons in the hadronic interpolating field and the factor
in the denominator arises because all octet baryons have two quarks of
the same flavour except from the $\Sigma^0$ and $\Lambda$.  This
algorithm can be efficiently implemented and is computationally
feasible for small systems, $A\alt 10$.  As an example of this method,
a $\null^4{\rm He}$ two point correlation function can be computed in
$\sim 0.8$ seconds per time slice on a single core of a Dual Core AMD
Opteron 285 processor.

\section{Multi-baryon contractions with determinants}

For larger atomic number, $A\agt 10$, alternative methods are required
to perform the contractions in a computationally feasible manner.  It
is straightforward to see how this can be done by examining the two
point functions above and making use of Wick's theorem
\cite{Wick:1950ee}.  The numerator of Eq.~\ref{eq:2pt} before the
integration over the gauge fields is performed is given by
\begin{eqnarray}
  \left[{\cal N}^h_1(t)\bar{\cal N}^{h}_2(0)\right]_{U} &=& \int {\cal D} q {\cal D} \bar q \;e^{-S_{QCD}[U]}\;
  \sum_{k'=1}^{N'_w}\sum_{k=1}^{N_w}
  \tilde{w}'^{(a'_1,a'_2\cdots a'_{n_q}),k'}_h\;\tilde{w}^{(a_1,a_2\cdots a_{n_q}),k}_h \times
  \nonumber\\
  && 
  \sum_{\bf j}  \sum_{\bf i}  \epsilon^{j_1,j_2,\cdots,j_{n_q}}\epsilon^{i_1,i_2,\cdots,i_{n_q}}
  {q}(a'_{j_{n_q}})\cdots q(a'_{j_2})q(a'_{j_1})
  \times
  \bar{q}(a_{i_1}) \bar q(a_{i_2})\cdots \bar q(a_{i_{n_q}})
  \,,
\end{eqnarray}
where the primed and unprimed indices are associated with the sink and
source interpolating fields, respectively and are composite colour,
spinor, flavour and spatial indices and $[\ldots]_U$ indicates the
value of the enclosed expression on a fixed gauge field.  The
Grassmann integral over quark fields can now be performed, resulting
in the replacement of the $q\overline q$ pairs by elements of the
quark propagator.
\begin{eqnarray}
  \left[{\cal N}^h_1(t)\bar{\cal N}^{h}_2(0)\right]_U &=&  \;e^{-{\cal S}_{eff}[U]}\;
  \sum_{k'=1}^{N'_w}\sum_{k=1}^{N_w}
  \tilde{w}'^{(a'_1,a'_2\cdots a'_{n_q}),k'}_h\;\tilde{w}^{(a_1,a_2\cdots a_{n_q}),k}_h \times
  \nonumber\\
  && 
  \sum_{\bf j}  \sum_{\bf i}  \epsilon^{j_1,j_2,\cdots,j_{n_q}}\epsilon^{i_1,i_2,\cdots,i_{n_q}}
  S(a'_{j_1};a_{i_1}) S(a'_{j_2};a_{i_2}) \cdots S(a'_{j_{n_q}};a_{i_{n_q}})
  \,,
  \label{eq:props}
\end{eqnarray}
where $S_{eff}[U]$ denotes the pure gauge part of the QCD action
together with the logarithm of the determinant of the Dirac matrix.
The above expression of Wick's theorem, can be written in terms of the
determinant of a matrix $G$ whose matrix elements are given by
\begin{equation}
  G({\bf a}^\prime;{\bf a})_{j,i}= \left\{ \begin{array}{ll} S(a'_{j};a_{i}) & {\rm for } \;\; a'_j \in {\bf a}'\;\; {\rm and}\;\; a_i \in {\bf a} \\
      \delta_{a'_j,a_i} & {\rm otherwise} \end{array}\right.\,,
\end{equation}
where, as before, ${\bf a}'=(a'_1,a'_2\cdots a'_{n_q})$ and ${\bf
  a}=(a_1,a_2\cdots a_{n_q})$. Note also that the non-trivial block
of the matrix $ G({\bf a}^\prime;{\bf a})$ is of size $n_q\times n_q$,
hence for computing its determinant we only need to consider this block. For this reason, in the following discussion,
the matrix $ G({\bf a}^\prime;{\bf a})$ denotes only this
small non-trivial block.

Making use of this definition, the full nuclear correlation function
can be written as
\begin{equation}
  \langle{\cal N}^h_1(t)\bar{\cal N}^{h}_2(0)\rangle = \frac{1}{\cal Z} \int {\cal D} {\cal U}\; e^{-{\cal S}_{eff}}\;
  \sum_{k'=1}^{N'_w}\sum_{k=1}^{N_w}
  \tilde{w}'^{(a'_1,a'_2\cdots a'_{n_q}),k'}_h\;\tilde{w}^{(a_1,a_2\cdots a_{n_q}),k}_h \times 
  \det{G({\bf a}';{\bf a})}\,.
  \label{eq:det}
\end{equation}
The determinant of a matrix of size $n_q$ can be evaluated in $n_q^3$
operations (for example via LU decomposition) instead of the naive $n_q!$ operations,
so making use of this representation of the nuclear correlation
function is numerically advantageous. Furthermore, because of the
flavour-blindness of the strong interaction, the matrix
$ G({\bf a}^\prime;{\bf a})$ is block
diagonal, as a result the determinant calculation breaks into a product of
smaller determinants, one for each flavour. 

Given the reduced weights determined above and appropriate quark
propagators, the implementation of Eq.~\ref{eq:det} is very fast,
scaling polynomially with the number of terms in the source and sink
quark level interpolating fields as well as the number of quarks per
flavour.
\begin{figure}[!t]
  \centering
  \includegraphics[width=6cm]{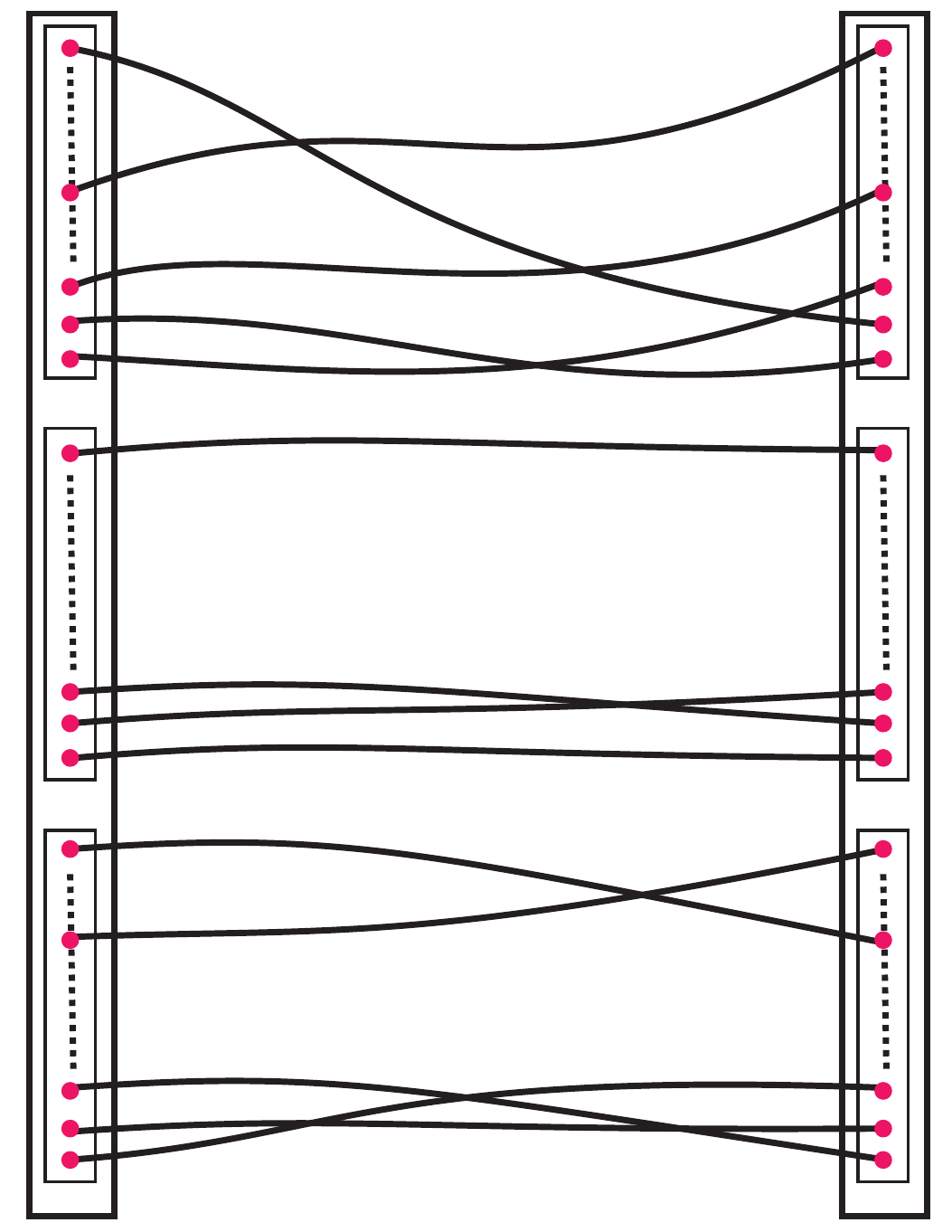}
  \caption{Illustration of the quark determinant level contraction
    with each of the three sub-blocks listing the up, down and strange
    quarks of the wave-function term respectively . Within each
    flavour of quark, all possible contractions are performed by
    forming a determinant of the matrix of quark propagators as
    described in the text.}
  \label{fig:q-q}
\end{figure}
The total cost of this form of contractions scales as
\begin{equation}
  n_u^3 n_d^3 n_s^3 \times N_w' N_w
\end{equation}
where $N_w'$ $N_w$ are the number of terms in the sink and source
quark interpolating fields respectively.\footnote{The expectation of polynomial scaling of 
contractions  was noted by D.~B.~ Kaplan in Ref.~\cite{OFT}. However, the scaling of $N_w$ and $N_w'$ with the atomic number 
$ A$
is generically exponential.} As a result, if we can construct interpolating fields with sufficiently small number of terms, correlation functions with a very large atomic number $A$ can be constructed.

\section{Nuclear correlation functions}
We have performed preliminary studies to
investigate the numerical efficiency of these methods.
Results for the quark-hadron approach have been presented in Ref.~\cite{NPLQCD_hyper} and here we focus on the determinant-based approach.  
 Calculations are performed on
an ensemble of gauge configurations generated with a tadpole-improved
L\"uscher-Weisz gauge action and a clover fermion action with
tadpole-improved tree-level clover coefficient. The gauge links
entering the fermion action are stout smeared, with $\rho=0.125$.
Three flavours of quarks with masses corresponding to the physical strange quark
mass were used.  The lattice spacing, $a \sim 0.145\ {\rm fm}$, and
the dimensions of the lattice are $L^3\times T = 32^3\times 48$
corresponding to a physical volume of $(4.6\ {\rm fm})^3\times 7.0\
{\rm fm}$ (further details will be presented elsewhere \cite{ISO}).
We have performed a large number of measurements from spatially
distinct sources on an ensemble of a ${\cal O}(250)$ gauge
configurations well separated in HMC evolution time. All calculations
are performed in double precision and care is taken to preserve the
dynamic range of correlation functions by rescaling quark propagators
before contractions are performed.

\begin{figure}[t]
  \centering
  \includegraphics[width=0.77\textwidth]{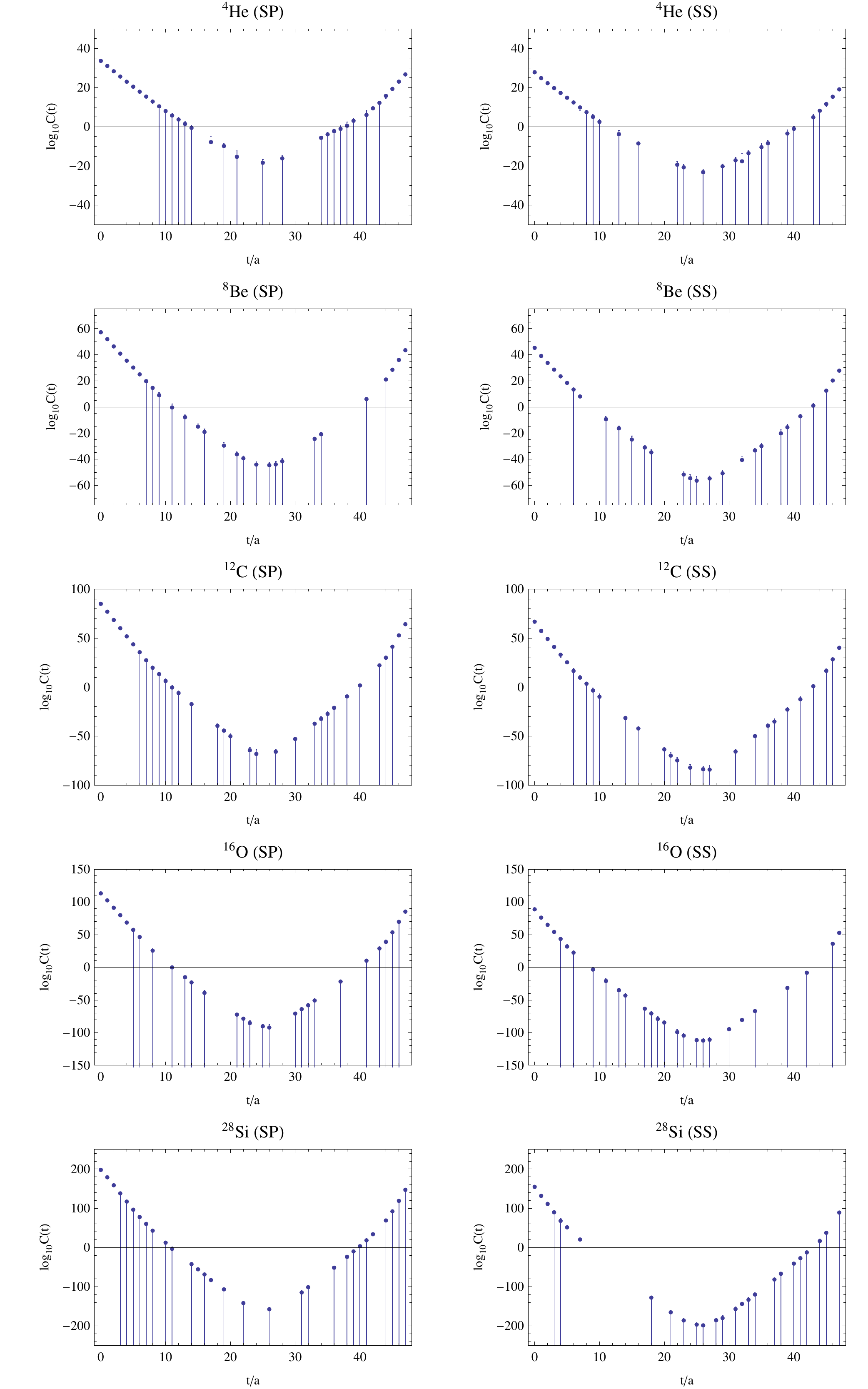}
  \caption{Correlation functions for nuclear systems, $^4$He, $^8$Be,
    $^{12}$C, $^{16}$O and $^{28}$Si. In each row, correlators
    based on both smeared-point (SP) and smeared-smeared (SS) quark
    propagators are shown.}
  \label{fig:nuke}
\end{figure}

In Fig.~\ref{fig:nuke}, the logarithms of correlation functions are
shown for correlators with the quantum numbers of the light
nuclei, $^4$He, $^8$Be, $^{12}$C, $^{16}$O and $^{28}$Si. Error bars
that reach the lower axis of the plots indicate that the correlator
has fluctuations that are negative at one standard deviation.  The
extracted energies for each of the atomic number $A<20$ systems are
consistent with a system of $A$ nucleons but with large uncertainties
at present (for $^{28}$Si, no flattening of the effective mass is seen
before the signal is lost).  Given the large number of near-threshold
energy levels expected in these complex nuclear systems (see
Ref.~\cite{NPLQCD_hyper} for an example for $A=4$), a clean extraction
of the ground state binding energies of these systems is beyond the
current work. In addition, the baryon number density of the larger
systems ($0.3\ {\rm fm}^{-3}$ for ${}^{28}{\rm Si}$) is substantial
and volume effects are expected to be significant. It will be
necessary to use larger volumes, increase greatly the statistical
precision and improve the interpolating operators that we have used in
order to obtain the binding energies and excitation spectra of these
systems. Nevertheless, this study demonstrates the computational
feasibility of lattice QCD calculations of light nuclei.

\section{Conclusions and outlook}

In this work, have discussed a systematic way of constructing
interpolating fields for multi-baryon systems. In addition, we have
investigated the issue of performing the Wick contractions with these
interpolating fields in lattice QCD (see
Ref.~\cite{Doi:2012xd} for related work). We have shown that there are
approaches that enable calculations of systems with very large number
of nuclei that are computationally feasible and demonstrated their
effectiveness by calculating correlators with baryon number up to
$A=28$. Given the expected spectra of such complex systems,
significant advances are required in order to extract ground state
energies from these correlators. Finally, the methods described here may prove
useful in calculations of QCD at non-zero baryon density, where projection on  to
a given baryon number is required. 

\acknowledgments 
We thank M. G. Endres, D.~B.~Kaplan, M.~J.~Savage and the members of NPLQCD collaboration
for insightful
discussions on the topic of this work. We also thank R. Edwards and B. Jo\'{o}
for help with QDP++ and Chroma software suites~\cite{Edwards:2004sx} which are the
software basis of all computations presented here.
We acknowledge computational support from the National Energy Research
Scientific Computing Center (NERSC, Office of Science of the US DOE,
DE-AC02-05CH11231), and the NSF through XSEDE resources provided by
NICS.  This work was supported in part by DOE grants DE-AC05-06OR23177
(JSA) and DE-FG02-04ER41302.  WD was also supported by DOE OJI grant
DE-SC0001784 and Jeffress Memorial Trust, grant J-968.

\bibliography{baryon_contractions}{}

\bibliographystyle{apsrev}

\end{document}